\documentclass[print,onecolumn,showpaces,aps,showpacs]{revtex4}%
\usepackage{amsfonts}
\usepackage{amsmath}
\usepackage{amssymb}
\usepackage{graphicx}%
\providecommand{\U}[1]{\protect\rule{.1in}{.1in}}
\newtheorem{theorem}{Theorem}
\newtheorem{acknowledgement}[theorem]{Acknowledgement}

\begin{document}
\title{Optical Solitons in $\mathcal{PT}$-symmetric Potentials with Competing Cubic-Quintic Nonlinearity: Existence, Stability, and Dynamics}
\author{Pengfei Li$^{1*}$}

\author{Lu Li$^{2}$}
\author{Dumitru Mihalache$^{3}$}

\affiliation{$^{1}$Department of Physics, Taiyuan Normal University, Taiyuan, 030031, China}
\affiliation{$^{2}$Institute of Theoretical Physics, Shanxi University, Taiyuan 030006, China}
\affiliation{$^{3}$Horia Hulubei National Institute of Physics and Nuclear Engineering, Magurele, Bucharest, RO-077125, Romania}

\begin{abstract}
We address the properties of optical solitons that form in media with competing cubic-quintic nonlinearity and parity-time ($\mathcal{PT}$)-symmetric complex-valued external potentials. The model describes the propagation of solitons in nonlinear optical waveguides with balanced gain and loss. We study the existence, stability, and robustness of fundamental, dipole, and multipole stationary  solutions in this $\mathcal{PT}$-symmetric system. The corresponding eigenvalue spectra diagrams for fundamental, dipole, tripole, and quadrupole solitons are presented. We show that the eigenvalue spectra diagrams for fundamental and dipole solitons merge at a coalescence point $W_{c1}$, whereas the corresponding diagrams for tripole and quadrupole solitons merge at a larger coalescence point $W_{c2}$. Beyond these two merging points, i.e.,  when the gain-loss strength parameter $W_{0}$ exceeds the corresponding coalescence points, the eigenvalue spectra cease to exist. The stability of the stationary solutions is investigated by performing the linear stability analysis and the robustness to propagation of these stationary solutions is checked by using direct numerical simulations.
\end{abstract}
\email{lpf281888@gmail.com}
\maketitle

\section{Introduction}

It is a well known fact that loss is ubiquitous in physical media and is considered as a detrimental
factor. However, in parity-time ($\mathcal{PT}$)-symmetric physical systems, it may play a positive role and help to create stationary
localized solutions, as demonstrated twenty years ago in a seminal work by Bender and Boettcher in the
quantum mechanics framework \cite{Bender1}. In this setting, non-Hermitian Hamiltonians with $\mathcal{PT}%
$-symmetry can have entirely real eigenvalue spectra \cite{Bender2}. However, there
exists a necessary but not sufficient condition for such non-Hermitian Hamiltonians, namely the external
complex-valued potential $U(x)$ is requested to satisfy $U(x)=U^{\ast}(-x)$,
where the asterisk stands for complex conjugation \cite{Bender3}.

The concept of $\mathcal{PT}$-symmetry has been introduced in other areas far beyond the quantum physics, such as optics and photonics \cite{OL32-2632} and Bose-Einstein condensates \cite{PRL101-080402,RRP_review}.  In optics, there is a growing interest in $\mathcal{PT}$-symmetric systems because the complex-valued external potential can be studied theoretically
\cite{OL32-2632,JPA38} and implemented experimentally
\cite{PRL103-093902,NP6,NP6b} in a series of physically-relevant settings. In the optics context the real part of the complex-valued potential stands for
the spatial distribution of the refractive index and the imaginary part stands for the balanced gain and loss in the corresponding optical waveguide structure. One of the key properties of a linear $\mathcal{PT}%
$-symmetric optical system is that there exists a certain threshold of gain-loss
strength \cite{PRL100-030402,PRL100-103904,PRA81-063807}. Below this threshold, the system
possesses all real eigenvalues (in the so-called the $\mathcal{PT}$-symmetric phase), but
the eigenvalues become complex (in the so-called $\mathcal{PT}$-symmetric broken phase) once
the gain-loss strength exceeds the above mentioned threshold. Moreover, the beam dynamics in $\mathcal{PT}$-symmetric optical systems exhibits some counterintuitive
characteristics, such as non-reciprocal light propagation, power
oscillations, optical transparency etc. During the past years $\mathcal{PT}$-symmetric optical
systems in nonlinear regimes have been investigated
extensively, and the key properties of diverse types of optical solitons, such as bright solitons, gap
solitons, Bragg solitons, gray or dark solitons and vortices that are supported by
various complex $\mathcal{PT}$-symmetric external potentials have been found
\cite{PRA83-041805,HYJ,ZX,HW1,HW2,PRA87-045803,YJK,PRA85-043840,PRA86-013808,Miri-PRA,RJP59,YJK1,Alexeeva2012, Igor1, Igor2, Li2015, RJP2016-review, Li2016b, Kartashov2016Optica, Mihalache2017, Lombard, Boris2017, Zhou2017, Yan2017}.
We mention here that two comprehensive reviews on the unique nonlinear features of a variety of $\mathcal{PT}$-symmetric physical systems have been recently published by Konotop {\it et al.} \cite{Konotop2016} and Suchkov {\it et al.} \cite{Suchkov2016}.

The generic model for the light propagation in nonlinear self-focusing Kerr-type media is the cubic nonlinear Schr\"{o}dinger (NLS) equation. However, for large light intensities it is absolutely necessary to
account for higher-order optical nonlinearities, especially, the quintic nonlinearity. Thus, the cubic-quintic (CQ) NLS equation has been proposed and studied extensively, and the obtained results show that the beam collapse can be arrested
by the defocusing quintic nonlinearity \cite{PRE61-3170,PRL88-073902,PRE74-066614,M1,M2}. Considering the effect of gain and loss, $\mathcal{PT}$-symmetric configurations with CQ nonlinearity, for example, two- and triple-core $\mathcal{PT}$-symmetric waveguides \cite{PRE88-062904,Chaos26-113103,PRE92-062909}, $\mathcal{PT}$-symmetric optical lattices \cite{OC285-1934,PRE91-023203}, and $\mathcal{PT}$-symmetric external potentials with competing nonlinearities \cite{PLA376-2880,OC354-277}, have been investigated in detail during the past years. In the present paper, based on $\mathcal{PT}$-symmetric CQ NLS equation, we explore the existence, stability, and robustness of different types of optical solitons and analyze the corresponding eigenvalue spectra by varying the gain-loss strength parameter. The stability and robustness of fundamental, dipole, and multipole solitonary solutions are systematically investigated by performing the corresponding linear stability analysis and by direct numerical simulations.

The paper is organized as follows. In the next Section, the generic NLS equation with competing cubic-quintic nonlinearities and its reductions are introduced. In Sec. 3, fundamental, dipole, tripole, and quadrupole stationary solutions are obtained by using adequate numerical methods. The influence of the gain-loss strength parameter on the $\mathcal{PT}$-symmetric eigenvalue spectrum diagrams is analyzed. In Sec. 4, we investigate systematically the stability and the dynamics of the stationary solutions. Finally, the conclusions are summarized in Sec. 5.

\section{Model and its reductions}

We begin our analysis by considering optical wave propagation in a planar
graded-index waveguide with cubic-quintic nonlinearity, which can be governed
by the following (1+1)-dimensional paraxial wave equation
\begin{equation}
i\frac{\partial A}{\partial z}+\frac{1}{2k_{0}}\frac{\partial^{2}
A}{\partial x^{2}}+\frac{k_{0}\left[  F(x)-n_{0}\right]  }{n_{0}}
A\label{Eq1}+\frac{k_{0}}{n_{0}}n_{2}\left\vert A\right\vert ^{2}A+\frac{k_{0}}{n_{0}
}n_{4}\left\vert A\right\vert ^{4}A=0\text{,}
\end{equation}
where $A(z,x)$ is the optical
field envelope function, $k_{0}=2\pi n_{0}/\lambda$ is the wavenumber with
$\lambda$ and $n_{0}$ being the wavelength of the optical source and the
background refractive index, respectively. Here, $F(x)=F_{R}(x)+iF_{I}(x)$ is a
complex-valued function, in which the real part represents the linear refractive
index distribution and the imaginary part stands for the gain and loss; $n_2$ and $n_4$ are the cubic and quintic nonlinear parameters, respectively.
Introducing the transformations $\Psi(\zeta,\xi)=\sqrt{\left\vert
n_{4}\right\vert /\left\vert n_{2}\right\vert }A(z,x)$, $\xi=k_{0}n_{2}%
\sqrt{2/(n_{0}\left\vert n_{4}\right\vert )}x$, and $\zeta=k_{0}n_{2}%
^{2}/(n_{0}\left\vert n_{4}\right\vert )z$, Eq. (\ref{Eq1}) can
be rewritten in a dimensionless form%
\begin{equation}
i\frac{\partial\Psi}{\partial\zeta}+\frac{\partial^{2}\Psi}{\partial \xi^{2}%
}+U(\xi)\Psi+\sigma_{1}\left\vert \Psi\right\vert ^{2}\Psi+\sigma
_{2}\left\vert \Psi\right\vert ^{4}\Psi=0\text{.} \label{Eq2}%
\end{equation}
Here $\sigma_{1}=n_{2}/\left\vert n_{2}\right\vert =\pm 1$ and $\sigma_{2}%
=n_{4}/\left\vert n_{4}\right\vert =\pm 1$, where $\pm 1$ corresponds to self-focusing ($+$) or
self-defocusing ($-$) situations,
respectively. The normalized potential is $U(\xi)\equiv V(\xi)+iW(\xi)$ with $V(\xi)=\left\vert
n_{4}\right\vert [F_{R}(x)-n_{0}]/n_{2}^{2}$ and $W(\xi)=\left\vert
n_{4}\right\vert F_{I}(x)/n_{2}^{2}$, which are required to be even and odd
functions, respectively, for  $\mathcal{PT}$-symmetric nonlinear optical waveguides.

We search for the stationary solutions of Eq. (\ref{Eq2}) in the form $\Psi(\zeta,\xi)=\phi(\xi
)e^{i\beta\zeta}$, where $\phi(\xi)$ is a complex-valued function and $\beta$ is the corresponding
propagation constant. Substitution into Eq. (\ref{Eq2}) yields%
\begin{equation}
\frac{d^{2}\phi(\xi)}{d\xi^{2}}+U\left(  \xi\right)  \phi(\xi)+\sigma_1
\left\vert \phi(\xi)\right\vert ^{2}\phi(\xi)+\sigma_2
\left\vert \phi(\xi)\right\vert ^{4}\phi(\xi)=\beta\phi(\xi)\text{.}
\label{sta_eq}%
\end{equation}
Here, we take the external potential as a super Gaussian-type function in the form%
\begin{equation}
V(\xi)=V_{0}e^{-\left(  \frac{\xi}{\xi_{0}}\right)  ^{2m}}\text{, \ }%
W(\xi)=W_{0}\left(  \frac{\xi}{\xi_{0}}\right)  e^{-\left(  \frac{\xi}{\xi
_{0}}\right)  ^{2m}}\text{,} \label{Potential}%
\end{equation}
where the parameters $V_{0}$ and $W_{0}$ are the normalized modulation
strengths of the refractive index and the balanced gain and loss, respectively, in which the parameter $W_{0}$
characterizes the degree of non-Hermiticity for the $\mathcal{PT}$-symmetric
system, $\xi_{0}$ is the width of the potential, and $m$ is the power index of the
super-Gaussian function. As a generic example, the power index is taken as $m=2$ in
this work, whereas the profile of $V(\xi)$ tends gradually to a rectangular
distribution with the increase of the integer parameter $m$.

\section{Stationary solutions and nonlinear eigenvalue spectrum diagrams}

In this Section, we explore the existence of stationary solutions of Eq. (\ref{Eq2}) and we obtain the nonlinear eigenvalue spectrum diagrams. The stationary solutions of Eq. (\ref{Eq2}) and their corresponding eigenvalues, i.e., the propagation constants, can be obtained by
solving numerically Eq. (\ref{sta_eq}). Here, for simplicity, we only consider the case of $\sigma_1=1$ and $\sigma_2=-1$.

It is well known that, in the absence of the nonlinear parameters, i.e., $\sigma_1=\sigma_2=0$, the eigenvalues become complex-valued with increasing of the gain-loss strength parameter, which leads to amplification or attenuation of optical fields during propagation. This indicates that the system undergos a ``phase transition" from a $\mathcal{PT}$-symmetric phase to a $\mathcal{PT}$-symmetry broken phase. However, this ``phase transition" is much different in the presence of the nonlinear cubic parameter, i.e., when $\sigma_{1}=1$ and $\sigma_{2}=0$. The obtained results show that the nonlinear eigenvalues can also become complex-valued from real ones with increasing of the gain-loss strength parameter, but the corresponding eigenvalue spectra undergo two such ``phase transitions". The first one is a bifurcation point from which the eigenvalue spectrum of the ground state is bifurcated into two branches, the real branch and the complex branch (note that the corresponding solutions have no physical relevance at the complex branch \cite{PRA86-013612,Fortscher}). The second one is a coalescence point for two modes, at which the two modes are terminated. Especially, the coalescence point as a function of the input power undergoes also a transition from the coalescence of the ground mode and the first excited mode to the coalescence of the first excited mode and the second excited mode \cite{RJP61-577}.

Here, we discuss the more general case in the presence of both cubic and quintic nonlinearities.
Similarly, when the eigenvalues for Eq. (\ref{sta_eq}) are real, the corresponding eigenstates are stationary solutions or nonlinear modes for the system (\ref{Eq2}). However, when the eigenvalues are complex-valued, the solutions for Eq. (\ref{sta_eq}) can only present the onset of optical fields and have no physical relevance, as shown in Refs. \cite{PRA86-013612,Fortscher,RJP61-577}. Thus, in this paper we only focus on the case of real eigenvalues.

\begin{figure}[th]
\centering\vspace{-0.0cm} \includegraphics[width=11.0cm]{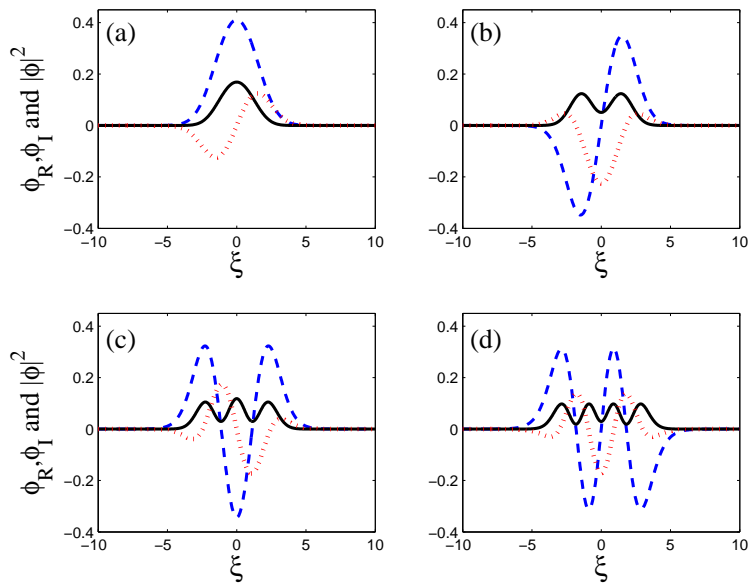}
\vspace{-0.0cm}\caption{(Color online) $\mathcal{PT}$-symmetric stationary solutions for Eq. (\ref{Eq2}). (a) Fundamental solution, (b) dipole solution, (c) tripole solution, (d) quadrupole solution, where the blue dashed and the red short-dashed curves are for real and imaginary parts of stationary solutions, respectively, and the dark solid curves represent the intensity
distributions. Here, the parameters are $V_{0}=6$, $\xi_{0}=4$, and $W_{0}=1$, at input power $P_{0}=0.5$.}%
\label{fig1}%
\end{figure}

Our numerical results confirm that the stationary solutions of Eq. (\ref{Eq2}) do exist. Figure \ref{fig1} presents the profiles of fundamental, dipole, tripole, and quadrupole stationary solutions, where the corresponding propagation constants are $5.72$, $5.05$, $4.08$, and $3.02$, respectively, and the system's parameters are chosen as $V_{0}=6$, $\xi_{0}=4$, and $W_{0}=1$, at the input power $P_{0}=0.5$.

\begin{figure}[th]
\centering\vspace{0.3cm} \includegraphics[width=9cm]{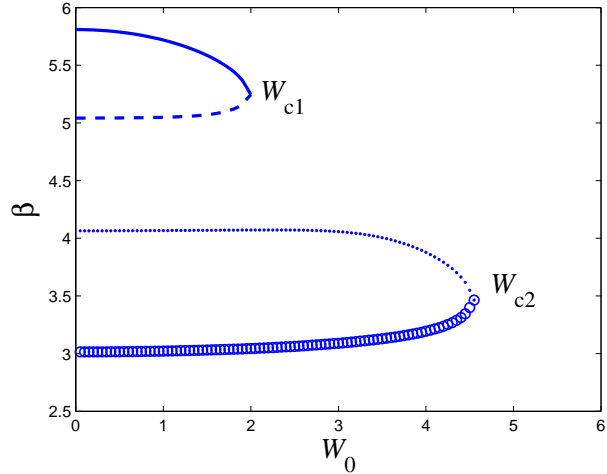}
\vspace{0.0cm}\caption{(Color online) The dependence of the propagation constant
$\beta$ on the gain-loss strength $W_{0}$ for input power $P_{0}=0.5$.
Here, the solid and dashed curves are eigenvalue spectra of fundamental and dipole
solutions, respectively, where the coalescence point is at $W_{c1}=2$. The dotted curve and circles
correspond to eigenvalue spectra of tripole and quadrupole solutions, respectively,  where the
coalescence point is at $W_{c2}=4.55$. Here the parameters are the same as in Fig. \ref{fig1}.}%
\label{fig2}%
\end{figure}

The eigenvalue spectrum diagrams, i.e., the dependence of the propagation constant on the gain-loss strength parameter for Eq. (\ref{Eq2}), are
presented in Fig. \ref{fig2}. From it, one can see that the eigenvalue spectra of the fundamental and dipole solutions  coalesce at the point $W_{c1}$. A similar behavior is also found for the eigenvalue spectra of tripole and quadrupole solutions displaying the second coalescence point $W_{c2}$, where $W_{c2}>W_{c1}$. It should be pointed out that at the coalescence points the eigenstates are degenerate except for the phase factor
$e^{i\pi/2}$.

\section{Stability analysis and propagations}

\begin{figure}[th]
\centering\vspace{0.0cm} \includegraphics[width=10cm]{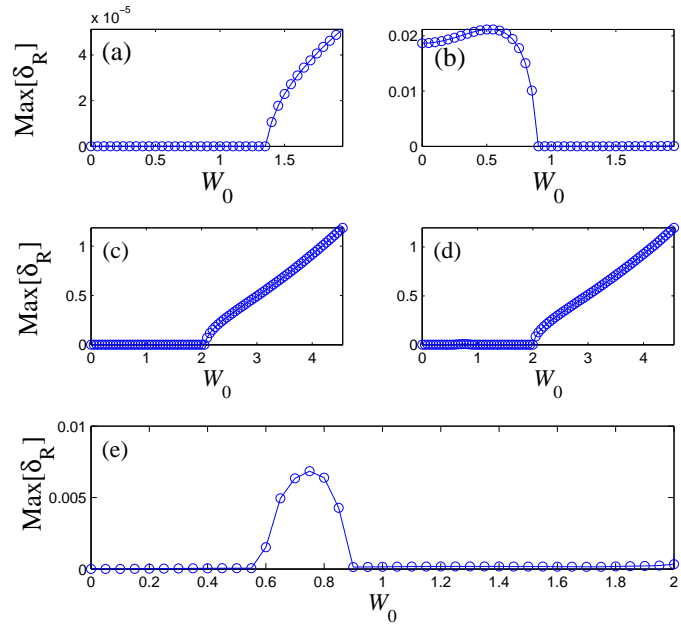}
\vspace{0.0cm}\caption{(Color online) The dependence of the largest real part of the eigenvalues for Eq.
(\ref{LSA1}) on the gain-loss strength $W_{0}$, where (a), (b), (c),  and
(d) are the largest real parts of the eigenvalues for fundamental, dipole,
tripole, and quadrupole solutions in the super Gaussian-type potential with
$m=2$ and $P_{0}=0.5$, respectively, and (e) the details of (d) in the region of $0\leq W_{0}\leq2$. Here the parameters are the same as in Fig. \ref{fig1}.}%
\label{fig3}%
\end{figure}

\begin{figure}[th]
\centering\vspace{0.0cm} \includegraphics[width=11cm]{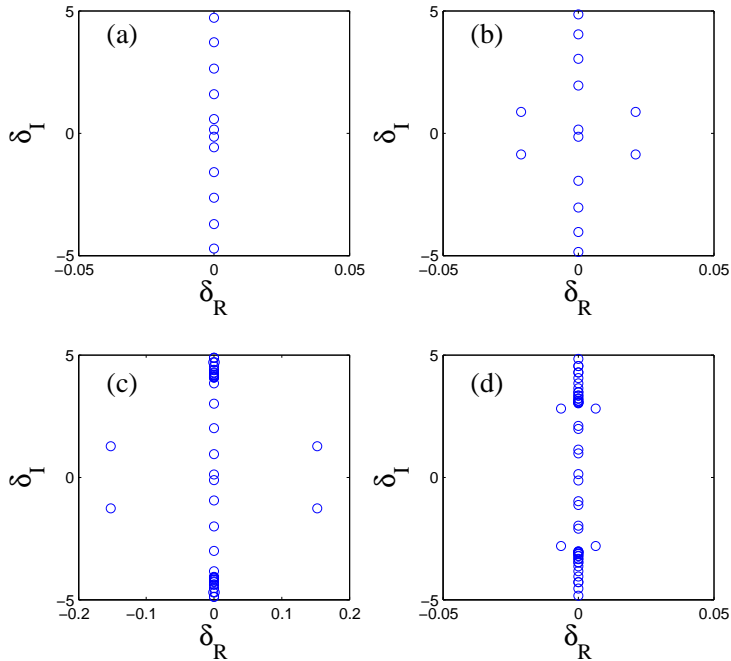}
\vspace{0.0cm}\caption{(Color online) The eigenvalue spectra of the stationary solutions. The real part $\delta_R$ and the imaginary part $\delta_I$ of the eigenvalues for the fundamental solution (a), the dipole solution (b), the tripole solution (c), and the quadrupole solution (d) with the gain-loss strength $W_{0}=1$, $0.5$, $2.2$, and $0.7$, respectively. Here, the other parameters are the same as in Fig. \ref{fig1}}%
\label{fig4}%
\end{figure}

\begin{figure}[th]
\centering\vspace{-0.0cm} \includegraphics[width=11cm]{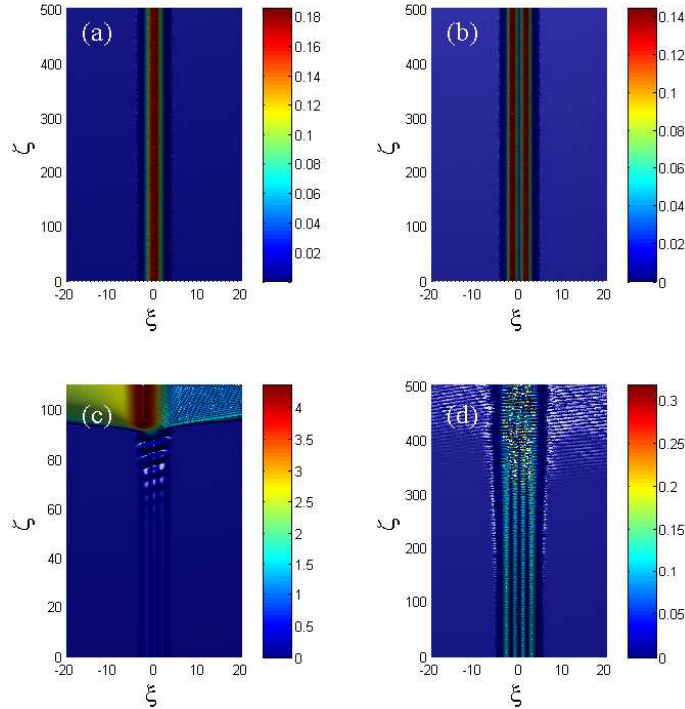}
\vspace{-0.3cm}\caption{(Color online) The evolution plots of the fundamental, dipole, tripole, and quadrupole solitons. The panels (a), (b), (c), and (d) correspond to the values of the gain-loss parameters $W_{0}$ as in Fig. \ref{fig4}. The other parameters are the same as in Fig. \ref{fig1}.}
\label{fig5}%
\end{figure}

In this Section, we will discuss the stability of the optical modes by
employing both the linear stability analysis and direct numerical simulations.

The linear stability analysis can be performed by adding a perturbation to a
known solution $\phi(\xi)$%
\begin{equation}
\Psi\left(  \xi,\zeta\right)  =e^{i\beta\zeta}\left[  \phi\left(  \xi\right)
+u\left(  \xi\right)  e^{\delta\zeta}+v^{\ast}\left(  \xi\right)
e^{\delta^{\ast}\zeta}\right]  \text{,} \label{Perturbation}%
\end{equation}
where $\phi(\xi)$ is the stationary solution with the real propagation constant
$\beta$, $u(\xi)$, and $v(\xi)$ are small perturbations with $\left\vert
u\right\vert $, $\left\vert v\right\vert \ll\left\vert \phi\right\vert $.
Substituting Eq. (\ref{Perturbation}) into Eq. (\ref{Eq2}) and keeping only
the linear terms, we obtain the following linear eigenvalue problem%
\begin{equation}
i\left(
\begin{array}
[c]{cc}%
L_{11} & L_{12}\\
L_{21} & L_{22}%
\end{array}
\right)  \left(
\begin{array}
[c]{c}%
u\\
v
\end{array}
\right)  =\delta\left(
\begin{array}
[c]{c}%
u\\
v
\end{array}
\right)  \text{,} \label{LSA1}%
\end{equation}
where $L_{11}=d^{2}/d\xi^{2}+U-\beta+2\sigma_{1}\left\vert \phi\right\vert
^{2}\allowbreak+3\sigma_{2}\left\vert \phi\right\vert ^{4}$, $L_{12}%
=\sigma_{1}\allowbreak\phi^{2}\allowbreak+2\sigma_{2}\allowbreak\phi
^{2}\left\vert \phi\right\vert ^{2}$, $L_{21}=-L_{12}^{\ast}$ and
$L_{22}=-L_{11}^{\ast}$, and $\delta$ is a complex eigenvalue. If $\delta$ contains
a real part, the solution $\phi(\xi)$ is linearly unstable, otherwise,
$\phi(\xi)$ is linearly stable. Thus, the real part of the complex eigenvalues can be used to measure the instability growth rate of the perturbation. In the following, the linear stability of the stationary solution is characterized by the largest real part of $\delta$.
Thus, if it is zero, the solution is linearly stable, otherwise, it is
linearly unstable.

In Fig. \ref{fig3} we show the dependence of the largest real part of the eigenvalues for Eq. (\ref{LSA1}) on the gain-loss strength $W_{0}$ at the input power $P_{0}=0.5$. From it, one can see that the largest real
eigenvalues of Eq. (\ref{LSA1}) are very close to zero for the fundamental solutions
in Fig. \ref{fig3}(a), thus the fundamental solutions are linearly stable. The dipole solutions are unstable in the interval of $0\leq W_{0}\leq 0.9$, but they are stable in the interval of $0.9\leq W_{0}\leq W_{c1}$, as shown in Fig. \ref{fig3}(b). The tripole solutions are linearly stable for $W_{0}$ ranging from $0$ to $2$, and they become unstable in the region of $2\leq W_{0}\leq W_{c2}$, see Fig. \ref{fig3}(c). For the quadrupole solutions, the unstable regions are $2\leq
W_{0}\leq W_{c2}$ and $0.55\leq W_{0}\leq0.9$, see Figs. \ref{fig3}(d) and \ref{fig3}(e).

In fact, in the linear stability analysis the eigenvalue spectra of Eq. (\ref{LSA1}) are quartet symmetric, i.e., the imaginary part $\delta_I$ and the real part $\delta_R$ of the eigenvalues appear in pairs for a $\mathcal{PT}$-symmetric system. As an example, we present in Fig. \ref{fig4} the eigenvalue spectra of the fundamental, dipole, tripole, and quadrupole solutions with the gain-loss strength parameters $W_{0}=1$, $0.5$, $2.2$, and $0.7$, respectively. One can see that the fundamental solution is stable, whereas the dipole, tripole, and quadrupole solutions are unstable for these particular values of the gain-loss strength parameter $W_{0}$. To confirm the results of the linear stability analysis, we performed the propagation of stationary solutions by numerically simulating Eq. (\ref{Eq2}), in which the fundamental, dipole, tripole, and quadrupole solitons are perturbed by a 5\% random noise. The corresponding soliton evolutions are summarized in Fig. \ref{fig5}. From it, it can be seen that the fundamental
solution can propagate robustly and the tripole and quadrupole solutions are unstable, see Figs. \ref{fig5}(a), \ref{fig5}(c),  and \ref{fig5}(d), respectively. For the dipole solution, the result of the linear stability analysis indicates that it is unstable, but the numerical simulation shows that it can propagate stably, as illustrated in Fig. \ref{fig5}(b). This is because the largest real part of the corresponding eigenvalue is very small, $\max(\delta_R)\approx0.0212$. This result shows  an example of a weak instability corresponding to a very small growth rate.

\section{Conclusions}

In summary, we have investigated the existence, stability, and robustness to perturbations of stationary solutions in a
competing cubic-quintic nonlinear optical waveguide with a $\mathcal{PT}$-symmetric super Gaussian-type external potential. We have reported the key properties of the fundamental,
dipole, tripole, and quadrupole solutions. Also, we have found that
the eigenvalue spectra of the fundamental and dipole solutions, as well as the tripole and quadrupole solutions merge at the coalescence points $W_{c1}$ and $W_{c2}$, respectively. Similar to the case of the self-focusing cubic NLS equation with a  $\mathcal{PT}$-symmetric potential, the eigenvalue spectra cease to exist when the gain-loss strength parameter $W_{0}$ exceeds the above mentioned coalescence points.
Finally, the stability of the fundamental, dipole, tripole, and quadrupole solutions
has been investigated by performing the linear stability analysis and has been checked by direct numerical simulations.

\begin{acknowledgement}
This research was supported by Doctoral Scientific Research Foundation of
Taiyuan Normal University No. I170144, by the National Natural Science
Foundation of China, through Grant No. 61475198, and by the
Shanxi Scholarship Council of China, through grant No. 2015-011.
\end{acknowledgement}

\end{document}